\documentclass[11pt,british]{article}
\usepackage[utf8]{inputenc}
\usepackage[a4paper]{geometry}
\usepackage{babel}

\usepackage{amstext}
\usepackage{esint}
\usepackage[unicode=true, 
 bookmarks=true,bookmarksnumbered=false,bookmarksopen=false,
 breaklinks=false,pdfborder={0 0 1},backref=false,colorlinks=false]
 {hyperref}
\hypersetup{pdftitle={Perturbations in the relaxation mechanism for a large cosmological constant},
 pdfauthor={Florian Bauer},
 pdfsubject={Perturbations and tracking in the model of arXiv:0902.2215},
 pdfkeywords={Perturbations, tracking, dark energy, cosmological constant, vacuum, relaxation, LXCDM, fine-tuning, coincidence problem}}
 
\makeatletter
\renewcommand\[{\begin{equation}}
\renewcommand\]{\end{equation}} 

\makeatother

\begin{document}

\title{Perturbations in the relaxation mechanism for a large cosmological
constant}

\author{\textbf{Florian Bauer}%
\thanks{fbauer@ecm.ub.es%
}\vspace*{0.5cm}
 \\
{\small High Energy Physics Group, Dept.\ ECM, and Institut de
Ciències del Cosmos}\\
{\small{} Universitat de Barcelona, Av.\ Diagonal 647, E-08028
Barcelona, Catalonia, Spain}}

\date{{ }}
\maketitle
\begin{abstract}
Recently, a mechanism for relaxing a large cosmological constant~(CC)
has been proposed \cite{Relax-LXCDM}, which permits solutions with
low Hubble rates at late times without fine-tuning. The setup is implemented
in the $\Lambda$XCDM framework, and we found a reasonable cosmological
background evolution similar to the $\Lambda$CDM model with a fine-tuned
CC. In this work we analyse analytically the perturbations in this
relaxation model, and we show that their evolution is also similar
to the $\Lambda$CDM model, especially in the matter era. Some tracking
properties of the vacuum energy are discussed, too.
\end{abstract}

\section{Introduction}

The cosmological constant problem has been around for a long time
\cite{CCProblem}. However, until the discovery of the accelerated
cosmic expansion \cite{Acceleration} there was the hope for a simple
symmetry mechanism which forces the CC to vanish. While such a mechanism
could still exist, it does not explain the cosmic acceleration, which
in the context of General Relativity requires a new dark energy component.
For that matter a tiny positive CC (equivalent to a tiny vacuum energy
density~$\rho_{\Lambda}$) represents the simplest solution. Alternative
candidates with more dynamics are, e.g., scalar field models~\cite{Scalars}
or modified gravity~\cite{ModGravity}. However, often a vanishing
(initial) CC is tacitly assumed, and many dark energy candidates provide
only for the current acceleration. The big problem of a vanishing
or tiny CC is of technical origin since the CC receives huge contributions
from phase transitions and quantum zero-point energy. They add up
to an enormous initial value~$|\rho_{\Lambda}^{i}|$ estimated in
the range $(10^{2}\cdots10^{19}\,\text{GeV})^{4}$, which is far above
the observed effective value~$\rho_{\Lambda}^{0}\sim(10^{-12}\,\text{GeV})^{4}$.
The usual method to {}``solve'' the problem is adding a counter-term~$\rho_{\Lambda}^{ct}$
such that $\rho_{\Lambda}^{i}+\rho_{\Lambda}^{ct}=\rho_{\Lambda}^{0}$.
Obviously,~$\rho_{\Lambda}^{ct}$ must be extremely fine-tuned otherwise
one would still end up with a large vacuum energy. A possible way
to avoid the fine-tuning is the introduction of a new mechanism or
energy component which dynamically relaxes the value of the CC. As
an example, it was shown in Ref.~\cite{Stefancic08} that a dark
energy component with an inhomogeneous equation of state is able to
achieve this task. Moreover, as the starting point of this work we
consider the recent CC relaxation model of Ref.~\cite{Relax-LXCDM}.
It has been constructed in the $\Lambda$XCDM framework~\cite{LXCDM,Pert-LXCDM-2},
where dark energy acts as a varying CC \cite{Overduin,Reuter,ShapiroSola,Babic,FBPhD,Klinkhamer08,Borges08,Urban09,LRun-Ward}.
A global fit to various models with a time-dependent CC has been performed
recently in Ref.~\cite{Pert-LXCDM-1}, and some of them turned out
to be perfectly compatible with the latest observational data. More
work related to CC relaxation can be found e.g.\ in Refs.~\cite{Barr:2006mp,Diakonos:2007au,Batra:2008cc}.

In the $\Lambda$XCDM relaxation model discussed here, the large initial
vacuum energy~$\rho_{\Lambda}^{i}$ is kept under control and the
CC relaxes dynamically without fine-tuning parameters. The resulting
cosmological evolution can be arranged to be similar to the $\Lambda$CDM
model with the usual ingredients like radiation and matter dominated
eras, and a final de~Sitter epoch. In this paper we provide a deeper
analysis of the relaxation mechanism and complement the original work
by analysing the evolution of perturbations. As a result, we will
find that they behave in many aspects very similar to those in $\Lambda$CDM.
Earlier work about perturbations in varying vacuum models can be found,
e.g.\ in Ref.~\cite{LRun-Pert-1}.

The paper is organised as follows: in Sec.~\ref{sec:Relaxation}
we briefly explain the CC relaxation mechanism and demonstrate the
absence of fine-tuning. In Sec.~\ref{sec:Tracking} we discuss the
tracking behaviour of the varying CC in the matter and radiation epochs.
This will be useful for the detailed analysis of perturbations in
Sec.~\ref{sec:Perturbations}. Finally, we present our conclusions
in Sec.~\ref{sec:Conclusions}.

\section{Relaxation mechanism\label{sec:Relaxation}}

Here, we repeat the basic principles of the CC relaxation model as
proposed in Ref.~\cite{Relax-LXCDM}. For concreteness, let us consider
a spatially flat universe with only dust matter, radiation and dark
energy. According to the Einstein equations, the Hubble expansion
rate~$H$ is related to the energy content by the Friedmann equation
\begin{equation}
H^{2}=\frac{8\pi G}{3}(\rho_{m}+\rho_{r}+\rho_{\Lambda}),\label{eq:Intro-Friedmann-1}\end{equation}
with~$G$ being Newton's constant. By~$H_{0}\sim10^{-42}\,\text{GeV}$
we denote today's Hubble rate, which is one of the smallest energy
scales in the universe. Now, assume for the moment that dark energy
was just a large constant CC term $\rho_{\Lambda}=\rho_{\Lambda}^{i}\gg\rho_{\Lambda}^{0}$.
Consequently, at some early time, this large value of $|\rho_{\Lambda}|$
would dominate over the matter and radiation energy densities $\rho_{m,r}$,
and would induce for $\rho_{\Lambda}^{i}>0$ a de~Sitter cosmos with
large~$H\gg H_{0}$, or respectively a Big Crunch for $\rho_{\Lambda}^{i}<0$.
Both solutions would prevent the well established Big Bang cosmology.
Since from a theoretical viewpoint a large $|\rho_{\Lambda}^{i}|$
seems unavoidable, some mechanism is needed to obtain back the standard
cosmic evolution. Recently, a possible solution has been presented
in Ref.~\cite{Relax-LXCDM} with the introduction of a dynamical
part into dark energy. In this model the total vacuum energy is given
by\[
\rho_{\Lambda}=\rho_{\Lambda}^{i}+\frac{\beta}{f},\,\,\,\beta=\text{const.}\]
which represents a varying CC term $\rho_{\Lambda}$ including the
(large) constant initial part~$\rho_{\Lambda}^{i}$ and a variable
function~$f$. This setup has been implemented in the $\Lambda$XCDM
framework \cite{LXCDM}, where the dark energy equation of state (EOS)
is always~$-1$, and an intrinsic interaction with another dark X
component is present. In our case, the X component is just pressureless
dark matter $\rho_{X}=\rho_{m}$, and its non-standard evolution is
determined by the Bianchi identities in the form of the modified conservation
equation\begin{equation}
\dot{\rho}_{\Lambda}+\dot{\rho}_{X}+3H\rho_{X}=0.\label{eq:L-X-Conservation}\end{equation}
The CC relaxation mechanism is working when the dynamical term~$\beta/f$
is compensating the huge constant part~$\rho_{\Lambda}^{i}$ in order
to obtain $|\rho_{\Lambda}|\ll|\rho_{\Lambda}^{i}|$. To start with
an example, remember that current observations suggest a universe
not far from a de~Sitter cosmos with a small Hubble rate~$H_{*}\approx H_{0}$.
Now we show that in this late-time epoch the function~$f=9H^{2}$
is sufficient for the CC relaxation \cite{Stefancic08}, and the relation
\begin{equation}
\rho_{\Lambda}=\rho_{\Lambda}^{i}+\frac{\beta}{9H^{2}}\approx\rho_{\Lambda}^{0}\label{eq:rhoL-simple-f}\end{equation}
can be realised by fixing the parameter~$\beta$ without fine-tuning:
\begin{equation}
\beta\approx-\rho_{\Lambda}^{i}9H_{*}^{2}.\label{eq:beta-relax-condition}\end{equation}
In the Friedmann equation~(\ref{eq:Intro-Friedmann-1}) at late times,\begin{equation}
\rho_{c}=\frac{3H^{2}}{8\pi G}=\rho_{\Lambda}^{i}+\frac{\beta}{9H^{2}}+\rho_{m},\label{eq:Friedmann-simple-f}\end{equation}
the terms $\rho_{\Lambda}^{i}$ and $\beta/(9H^{2})$ are dominating
over the other terms $\rho_{c}$ and $\rho_{m}$, because the initial
CC $\rho_{\Lambda}^{i}$ is expected to be much larger than the present
critical energy density $\rho_{c}^{0}=3H_{0}^{2}/(8\pi G)$, and $\beta/(9H^{2})$
is large because of the smallness of the late-time Hubble rate~$H\sim H_{0}$.
Therefore, we solve Eq.~(\ref{eq:Friedmann-simple-f}) for~$H$
and find in good approximation\begin{equation}
9H^{2}=\frac{-\beta}{\rho_{\Lambda}^{i}},\label{eq:H2proptoBeta}\end{equation}
where terms of the order $H^{2}\rho_{c}/\rho_{\Lambda}^{i}$ have
been neglected. In the last equation we observe that a tiny Hubble
rate~$H$ is suggested by the large value of~$\rho_{\Lambda}^{i}$
in the denominator, thus it is a consistent solution and not an assumption.
More important than the smallness of~$H$ is its dependence on the
parameter~$\beta$. It is clear that $\beta$ must have a certain
value as given in Eq.~(\ref{eq:beta-relax-condition}) for obtaining~$H\approx H_{*}$,
but no fine-tuning is necessary because of $H^{2}\propto\beta$ in
Eq.~(\ref{eq:H2proptoBeta}). As a result, a change of~$10\%$ in
the parameter~$\beta$ would change the phenomenology~($H^{2}$)
only by roughly~$10\%$. This demonstrates the absence of fine-tuning
in the CC relaxation mechanism, because small parameter variations
will induce only small changes in the resulting cosmology.

Let us compare the situation with the counter-term method, where for
concreteness the initial vacuum energy density $\rho_{\Lambda}^{i}=10^{60}\,\rho_{c}^{0}$
is taken to be much larger than the present critical energy density~$\rho_{c}^{0}$.
In this case the Friedmann equation reads \[
\rho_{c}=\frac{3H^{2}}{8\pi G}=\rho_{\Lambda}^{i}+\rho_{\Lambda}^{ct}+\rho_{m}.\]
For a universe with low Hubble rate~$H$ we need~$\rho_{c}\approx\rho_{c}^{0}$,
and the counter term~$\rho_{\Lambda}^{ct}$ must be chosen with extremely
high accuracy,\[
\rho_{\Lambda}^{ct}=\rho_{c}^{0}-\rho_{\Lambda}^{i}=-\rho_{\Lambda}^{i}\left(1-10^{-60}\right),\]
where we neglected~$\rho_{m}$ for simplicity. As above, imagine
a small change by roughly~$10\%$ in this fine-tuned counter-term,
$\rho_{\Lambda}^{ct}=-\frac{9}{10}\,\rho_{\Lambda}^{i}$, then we
would find\[
\rho_{c}=\frac{3H^{2}}{8\pi G}=\rho_{\Lambda}^{i}+\left(-\frac{9}{10}\rho_{\Lambda}^{i}\right)=\frac{1}{10}\rho_{\Lambda}^{i}=10^{59}\,\rho_{c}^{0}\gg\rho_{c}^{0}\]
and the resulting Hubble rate would be much larger than the present
value~$H_{0}$. The need for fine-tuning is obvious in this example. 

In contrast to this, the parameter~$\beta$ in the CC relaxation
mechanism does not suffer from this problem. Moreover, the corresponding
de~Sitter final regime is dynamically stable. To understand this,
note that the two big terms in $\rho_{\Lambda}$ in Eq.~(\ref{eq:rhoL-simple-f})
act in opposite directions. Consider the case $\rho_{\Lambda}^{i}<0,\,(\beta>0)$
and $\rho_{\Lambda}^{i}$ dominating over $\beta/(9H^{2})$. This
would induce through $\rho_{\Lambda}\approx\rho_{\Lambda}^{i}<0$
a decreasing $H$ until the positive value of $\beta/(9H^{2})$ becomes
large enough to compensate the negative $\rho_{\Lambda}^{i}$. On
the other hand, when $\beta/(9H^{2})>0$ is dominating over~$\rho_{\Lambda}^{i}$,
the Hubble rate would increase until the equilibrium with $\rho_{\Lambda}^{i}$
is achieved again. These arguments work analogously in the $\rho_{\Lambda}^{i}>0,\,(\beta<0)$
case. The crucial reason for the absence of fine-tuning is that the
equilibrium point $H_{*}$ is determined only by the two large terms
in $\rho_{\Lambda}$ and not by the small energy densities $\rho_{\Lambda}$,
$\rho_{m}$ or $\rho_{r}$. Therefore, there is no need for subtracting
by hand large numbers from each other to obtain a small number, this
happens dynamically in the way described above.

For relaxing the CC also in the matter and radiation eras, the function
$f$ should be proportional to $(q-\frac{1}{2})$ and $(q-1)$, respectively,
where $q=-\ddot{a}a/\dot{a}^{2}$ is the deceleration parameter. Since
in the matter era $q\approx\frac{1}{2}$ (radiation era: $q\approx1$),
it is the smallness of $(q-\frac{1}{2})$ (or $(q-1)$) which keeps
$\beta/f$ close to $(-\rho_{\Lambda}^{i})$ in a dynamical way such
that $|\rho_{\Lambda}|\ll|\rho_{\Lambda}^{i}|$.

To obtain a full cosmological evolution with radiation, matter and
final de~Sitter eras, the following function has been constructed:\begin{equation}
f=4H^{2}\frac{(\frac{1}{2}-q)(2-q)}{(1-q)}+y\cdot72H^{6}(1-q)(1+q^{2}).\label{eq:f-Def-H-q}\end{equation}
The term proportional to the constant~$y$ is responsible for the
CC relaxation in the radiation regime, whereas the first term does
the same in the matter and de~Sitter stages. This particular sequence
follows from the different powers of $H$ in $f$ since $H$ decreases
with cosmological time. Moreover, $y=H_{eq}^{-4}$, where $H_{eq}\sim10^{5}H_{0}$
is the Hubble scale at the radiation-matter transition.

The function~$f$ in Eq.~(\ref{eq:f-Def-H-q}) can be expressed
in terms of geometrical scalars $R,S,T$ such that $\rho_{\Lambda}$
is a covariant scalar quantity, too:\begin{equation}
\rho_{\Lambda}=\rho_{\Lambda}^{i}+\frac{\beta}{f}=\rho_{\Lambda}^{i}+\beta\frac{R}{B}\,\,\,\,\text{with}\,\,\,\, B:=R^{2}-S+y\cdot R^{2}T.\label{eq:rhoL-Def}\end{equation}
Here, the Ricci scalar $R:=g^{ab}R_{ab}=6H^{2}(1-q)$, $S:=R_{ab}R^{ab}=12H^{4}(q^{2}-q+1)$
and $T:=R_{abcd}R^{abcd}=12H^{4}(q^{2}+1)$ are constructed from the
metric $g_{ab}$, the Ricci tensor $R_{ab}$ and respectively from
the Riemann tensor $R_{abcd}$, which will be fully defined in Sec.~\ref{sub:Pert-Grav}.
Finally, we remark that $\rho_{\Lambda}$ in Eq.~(\ref{eq:rhoL-Def})
defines a phenomenological model in the $\Lambda$XCDM framework,
it should not be confused with models based on a modified action~\cite{ModGravity}.
The current framework avoids some problems (e.g.\ the Ostrogradski
instability) usually present when the terms~$S$ and $T$ appear
explicitely in the action. However, the relaxation mechanism in the
modified gravity approach will be discussed in the future.

In the following sections we will need the Friedmann equations for
$H$ and $q$,\begin{eqnarray}
H^{2} & = & \frac{H_{0}^{2}}{\rho_{c}^{0}}(\rho_{X}+\rho_{\Lambda}+\rho_{r}),\label{eq:Friedmann-H}\\
qH^{2} & = & \frac{H_{0}^{2}}{\rho_{c}^{0}}\left(\frac{1}{2}\rho_{X}-\rho_{\Lambda}+\rho_{r}\right),\label{eq:Friedmann-q}\end{eqnarray}
where~$\rho_{c}^{0}=3H_{0}^{2}/(8\pi G)$ is today's critical energy
density. For simplicity, we consider only energy components with constant
equations of state. In this sense, radiation denotes photons and massless
neutrinos, the X component plays the role of dust-like dark matter
without pressure, and vacuum energy has the EOS $(-1)$ of a CC, respectively.

\section{Tracking behaviour\label{sec:Tracking}}

In addition to a viable background evolution, the model in Eq.~(\ref{eq:rhoL-Def})
also features an interesting tracking behaviour of the varying vacuum
energy~$\rho_{\Lambda}$. We will now analyse this in more detail
and make use of the results when discussing the perturbations in Sec.~\ref{sec:Perturbations}.

\subsection{Tracking in the radiation era\label{sub:Rad-Tracking}}

From $\rho_{\Lambda}$ in Eq.~(\ref{eq:rhoL-Def}) we observe that
compensating the large~$\rho_{\Lambda}^{i}$ means $|\rho_{\Lambda}^{i}+\beta R/B|\ll|\rho_{\Lambda}^{i}|$.
Thus, the smallness of $|B|={\cal O}(\beta R/\rho_{\Lambda}^{i})\ll H^{4}$
is a useful condition for characterising the relaxation of the CC
in the following. Note that this relation also implies that $B$ does
not vanish. 

In the radiation era ($q\approx1$) the function~$B$ is well approximated
by \[
B=-12H^{4}+2y\cdot24H^{4}R^{2}=12H^{4}(2yR^{2}-1),\]
indicating $R\approx1/\sqrt{2y}\sim H_{eq}^{2}$ in the minimum of
$B$ as a result of $|B|\ll H^{4}$. Now, let us solve Eq.~(\ref{eq:rhoL-Def})
for $\rho_{\Lambda}$ with~$R=6H^{2}(1-q)$ given by the Friedmann
equations (\ref{eq:Friedmann-H},\ref{eq:Friedmann-q}) \[
R=\frac{6H_{0}^{2}}{\rho_{c}^{0}}\left(\frac{1}{2}\rho_{X}+2\rho_{\Lambda}\right).\]
First, in the relaxation regime, $R\approx1/\sqrt{2y}$ has been found
above, and the approximation \[
(2yR^{2}-1)\approx2(\sqrt{2y}R-1),\]
simplifies the relation\begin{equation}
\rho_{\Lambda}=\rho_{\Lambda}^{i}+\beta\frac{R}{B}=\rho_{\Lambda}^{i}+\frac{\gamma}{\epsilon\cdot(\frac{1}{2}\rho_{X}+2\rho_{\Lambda})-1},\label{eq:rL-rLi-rX-rL}\end{equation}
with\[
\gamma:=\frac{\beta/\sqrt{2y}}{12H^{4}\cdot2},\,\,\,\,\epsilon:=\frac{6H_{0}^{2}}{\rho_{c}^{0}}\sqrt{2y}\sim\frac{1}{\rho_{c}^{eq}},\]
where $\rho_{c}^{eq}=3H_{eq}^{2}/(8\pi G)$ is the critical energy
density at the time of the radiation-matter transition. Solving Eq.~(\ref{eq:rL-rLi-rX-rL})
we find two solutions for $\rho_{\Lambda}$:\[
\rho_{\pm}=\frac{1}{8}\left[4\rho_{\Lambda}^{i}-\rho_{X}+\frac{2}{\epsilon}\pm\left|4\rho_{\Lambda}^{i}+\rho_{X}-\frac{2}{\epsilon}\right|\sqrt{1+\frac{32\gamma}{\epsilon(4\rho_{\Lambda}^{i}+\rho_{X}-\frac{2}{\epsilon})^{2}}}\,\right].\]
The square root can be expanded up to first order $\sqrt{1+x}\approx1+\frac{1}{2}x$
in the term $x=32\gamma/\cdots$ and we obtain\[
\rho_{\pm}\simeq\frac{1}{8}\left[4\rho_{\Lambda}^{i}-\rho_{X}+\frac{2}{\epsilon}\pm\left|4\rho_{\Lambda}^{i}+\rho_{X}-\frac{2}{\epsilon}\right|\pm8\kappa\right],\,\,\,\,\kappa:=\frac{2\gamma}{\epsilon|4\rho_{\Lambda}^{i}+\rho_{X}-\frac{2}{\epsilon}|}\sim\rho_{c}^{0}\left(\frac{H_{eq}}{H}\right)^{4}.\]
There are several cases to discuss. For $\rho_{\Lambda}^{i}<0$ we
find in the limit $\rho_{X}\ll|4\rho_{\Lambda}^{i}|$ a tracking solution\[
\rho_{+}\simeq-\frac{1}{4}\rho_{X}+\frac{1}{2\epsilon}+\kappa,\]
whereas for $\rho_{X}\gg|4\rho_{\Lambda}^{i}|$ the vacuum energy
density takes on its initial value $\rho_{\Lambda}\simeq\rho_{\Lambda}^{i}$
corresponding to a standard cosmos with constant $\rho_{\Lambda}$
at very high redshift, where $\rho_{\Lambda}^{i}$ is not dominant,
yet. We discard the $\rho_{-}$ solution since it does not have a
sensible behaviour for $\rho_{X}\ll|4\rho_{\Lambda}^{i}|$.

In the case $\rho_{\Lambda}^{i}>0$, we find the tracking solution
\[
\rho_{-}\simeq-\frac{1}{4}\rho_{X}+\frac{1}{2\epsilon}-\kappa,\]
for both limits $\rho_{X}\ll|4\rho_{\Lambda}^{i}|$ and $\rho_{X}\gg|4\rho_{\Lambda}^{i}|$.
Therefore, the tracking regime is persistent in the radiation era
and it will start right after inflation/reheating. Here, the second
branch $\rho_{+}$ is discarded for the same reasons as above.

Summarising, during the radiation epoch the CC relaxation involves
the tracking relation $\rho_{\Lambda}=-\frac{1}{4}\rho_{X}$ between
vacuum energy and dark/X matter, which implies via the conservation
equation~(\ref{eq:L-X-Conservation}) that all energy densities $\rho_{\Lambda}$,
$\rho_{X}$ and $\rho_{r}$ scale like radiation: $\dot{\rho}+4H\rho=0$.

Finally, we remark that the limits found here are the same as those
in Ref.~\cite{Relax-LXCDM}, where the simpler approximation $B\approx y\cdot R^{2}T$
was used.

\subsection{Tracking in the matter era}

Here, we show that the vacuum energy density~$\rho_{\Lambda}$ is
tracking the radiation energy density~$\rho_{r}$ in the matter era,
too. In this epoch, where $q\approx\frac{1}{2}$, we find\begin{equation}
\rho_{\Lambda}=\rho_{\Lambda}^{i}+\frac{\beta}{f}\approx\rho_{\Lambda}^{i}+\frac{\beta}{12H^{2}(\frac{1}{2}-q)+45H^{6}y}=\rho_{\Lambda}^{i}+\frac{1}{d(\frac{3}{2}\rho_{\Lambda}-\frac{1}{2}\rho_{r})+c}\label{eq:Mat-rL-Eq}\end{equation}
with\[
d:=\frac{12H_{0}^{2}}{\rho_{c}^{0}\beta}\,\,\,\text{and}\,\,\, c:=\frac{45H^{6}y}{\beta}.\]
Moreover, in the second step of Eq.~(\ref{eq:Mat-rL-Eq}) we replaced
$(\frac{1}{2}-q)$ by using the Friedmann equations~(\ref{eq:Friedmann-H},\ref{eq:Friedmann-q})
to obtain\[
H^{2}\left(\frac{1}{2}-q\right)=\frac{H_{0}^{2}}{\rho_{c}^{0}}\left(\frac{3}{2}\rho_{\Lambda}-\frac{1}{2}\rho_{r}\right),\]
which would be valid also for a finite baryon density $\rho_{b}\neq0$.
Solving Eq.~(\ref{eq:Mat-rL-Eq}) for $\rho_{\Lambda}$ yields\[
\rho_{\Lambda}=\frac{1}{6d}\left[x\pm\sqrt{x^{2}+12d(2+2c\rho_{\Lambda}^{i}-d\rho_{\Lambda}^{i}\rho_{r})}\right]\]
with $x:=-2c+3d\rho_{\Lambda}^{i}+d\rho_{r}$. After expanding the
root for large values of $x$ we find\[
\rho_{\Lambda}\approx\frac{1}{6d}\left[x\pm|x|\pm\frac{6d(2+2c\rho_{\Lambda}^{i}-d\rho_{\Lambda}^{i}\rho_{r})}{|x|}+\mathcal{O}(x^{-2})\right].\]
Since $d\rho_{\Lambda}^{i}=-\frac{4}{3}(\rho_{c}^{*})^{-1}<0$ the
{}``$+$'' solution corresponds to the relaxation regime ($|\rho_{\Lambda}|\ll|\rho_{\Lambda}^{i}|$).
The constant $\beta=-\rho_{\Lambda}^{i}9H_{*}^{2}$ is determined
in the final de~Sitter regime where~$H=H_{*}$ and~$\rho_{c}=\rho_{c}^{*}$,
cf.~Sec.~\ref{sec:Relaxation}. Finally, with $x\approx3d\rho_{\Lambda}^{i}=-4/\rho_{c}^{*}$
we obtain \begin{equation}
\rho_{\Lambda}\approx\frac{1}{2}\rho_{c}^{*}-\frac{2}{3}\frac{c}{d}+\frac{1}{3}\rho_{r}\approx\frac{1}{3}\rho_{r},\label{eq:Mat-3rL--rR}\end{equation}
where the first two terms can be neglected deep in the matter era
($\rho_{c}\approx\rho_{x}\propto a^{-3}$) because $\rho_{r}\gg\rho_{c}^{*}$
and\[
\frac{c}{d}=\frac{15}{4}\rho_{c}\left(\frac{\rho_{c}}{\rho_{c}^{eq}}\right)^{2}\sim\rho_{c}^{eq}\left(\frac{a}{a_{eq}}\right)^{-9}\ll\rho_{r}\sim\rho_{c}^{eq}\left(\frac{a}{a_{eq}}\right)^{-4},\]
respectively. As a result, the variable CC $\rho_{\Lambda}=\frac{1}{3}\rho_{r}$
is fixed, and tracks the radiation energy density. This is in contrast
to the tracking relation in the radiation era, $\rho_{r}\propto\rho_{\Lambda}=-\frac{1}{4}\rho_{X}$,
where the proportionality constant is not fixed. Finally, we integrate
the conservation equation~(\ref{eq:L-X-Conservation}) and find the
exact expression for the dark/X matter energy density,\begin{equation}
\rho_{X}=\rho_{X0}\, a^{-3}-\frac{4}{3}\rho_{r},\label{eq:Mat-rX-exact}\end{equation}
which shows that deviations from the standard dust scaling rule $\rho_{X}\propto a^{-3}$
are of the order~$\rho_{r}=\rho_{r0}a^{-4}$. Due to the interaction
between the vacuum and X components, the integration constant $\rho_{X0}$
is not exactly equal to the current X energy density, whereas $\rho_{r0}$
denotes the current radiation energy density.

\section{Perturbations\label{sec:Perturbations}}

We discuss the behaviour of linear perturbations in the relaxation
model and compare the results with the $\Lambda$CDM model \cite{ModernCosmo}.
Since the evolution equations for perturbations of photons, neutrinos
and baryons are not changed with respect to $\Lambda$CDM, there is
no need to discuss them in this work. Only during the radiation epoch
we consider the photon monopole and dipole modes.

\subsection{Perturbations in the gravitational sector\label{sub:Pert-Grav}}

In the following sections we analyse the scalar perturbations of the
metric in the Newtonian gauge, in which tensor and vector perturbations
are absent from the beginning. Additionally, it will be shown in Sec.~\ref{sub:Tensor-perturbations}
that the tensor modes (gravitational waves) in the $\Lambda$XCDM
relaxation model are unchanged in comparison to $\Lambda$CDM. We
use the convention $a,b,c,d,m,n=0\dots3$, $i,j,k=1\dots3$ for tensor
indices, where $x^{0}=t$ is the cosmological time and $\vec{x}=(x^{1},x^{2},x^{3})$
are spatial Euclidean coordinates. Commas denote partial derivatives
and semicolons covariant ones, in addition, the Einstein sum convention
is applied.

In linear order the perturbed metric reads\[
g_{00}=1+2\Psi(t,\vec{x}),\,\,\, g_{ij}=-\delta_{ij}a^{2}(t)(1+2\Phi(t,\vec{x})),\,\,\, g_{0i}=0,\]
where $a(t)$ is the cosmic scale factor and $\Phi$ and $\Psi$ are
the scalar potentials. In the Newtonian gauge, these quantities correspond
to the gauge-invariant Bardeen variables. With $g_{mn}$ the other
geometrical terms can be obtained in the usual way, where we use~$\frac{\ddot{a}}{a}=-H^{2}q$
and $H=\frac{\dot{a}}{a}$ frequently. Accordingly, we find the Christoffel
symbols\[
\Gamma_{00}^{0}=\dot{\Psi}\,\,\,\,\,\,\,\,\Gamma_{ij}^{0}=\delta_{ij}a^{2}\left(H(1+2\Phi-2\Psi)+\dot{\Phi}\right)\,\,\,\,\,\,\,\,\Gamma_{0i}^{0}=\Psi_{,i}\]
\[
\Gamma_{00}^{i}=a^{-2}\Psi_{,i}\,\,\,\,\,\,\,\,\Gamma_{jk}^{i}=\Phi_{,k}\delta_{ji}+\Phi_{,j}\delta_{ki}-\Phi_{,i}\delta_{jk}\,\,\,\,\,\,\,\,\Gamma_{j0}^{i}=\delta_{ij}\left(H+\dot{\Phi}\right),\]
from which the Riemann tensor\[
R_{\,\, bcd}^{a}=\Gamma_{bd,c}^{a}-\Gamma_{bc,d}^{a}+\Gamma_{mc}^{a}\Gamma_{bd}^{m}-\Gamma_{md}^{a}\Gamma_{bc}^{m}\]
and Ricci tensor $R_{bc}=R_{\,\, bca}^{a}$ are derived. The components
of the latter read \begin{eqnarray}
R_{00} & = & 3\frac{\ddot{a}}{a}+3\ddot{\Phi}+6H\dot{\Phi}-3H\dot{\Psi}-a^{-2}\nabla^{2}\Psi\\
R_{ik} & = & \Psi_{,ik}+\Phi_{,ik}-a^{2}\delta_{ik}\left[\ddot{\Phi}-a^{-2}\nabla^{2}\Phi+6H\dot{\Phi}-H\dot{\Psi}+\left(2-q\right)H^{2}(1+2\Phi-2\Psi)\right]\\
R_{0k} & = & 2\dot{\Phi}_{,k}-2H\Psi_{,k}.\end{eqnarray}
Furthermore, we obtain the Ricci scalar $R=g^{ab}R_{ab}$\begin{equation}
R=6H^{2}\left(1-q\right)+6\ddot{\Phi}+24H\dot{\Phi}-6H\dot{\Psi}-2a^{-2}\nabla^{2}\Psi-4a^{-2}\nabla^{2}\Phi-12H^{2}\left(1-q\right)\Psi,\label{eq:Pert-Scalar-R}\end{equation}
the squared Ricci tensor $S=R_{ab}R^{ab}$\begin{eqnarray}
S & = & 12H^{4}(q^{2}-q+1)+\ddot{\Phi}\left[12H^{2}\left(1-2q\right)\right]+\dot{\Phi}[72H^{3}(1-q)]+a^{-2}\nabla^{2}\Phi[-8H^{2}(2-q)]\nonumber \\
 &  & +a^{-2}\nabla^{2}\Psi\left[-4H^{2}\left(1-2q\right)\right]+\dot{\Psi}\left[-12H^{3}\left(1-2q\right)\right]+\Psi[-48H^{4}(q^{2}-q+1)],\label{eq:Pert-Scalar-S}\end{eqnarray}
and the squared Riemann tensor $T=R_{abcd}R^{abcd}$ \begin{eqnarray}
T & = & 12H^{4}(q^{2}+1)+\ddot{\Phi}[-24H^{2}q]+\dot{\Phi}[48H^{3}(1-q)]+\dot{\Psi}[24H^{3}q]\nonumber \\
 &  & +a^{-2}\nabla^{2}\Psi[8H^{2}q]+a^{-2}\nabla^{2}\Phi[-16H^{2}]+\Psi[-48H^{4}(q^{2}+1)].\label{eq:Pert-Scalar-T}\end{eqnarray}
The Gauß-Bonnet term $G=R^{2}-4S+T$ has the interesting property
that, in terms of $H$ and $q$, the scalar invariants $S$ and $T$
can be replaced by $S_{*}:=\frac{1}{3}R^{2}-\frac{1}{2}G$ and $T_{*}:=\frac{1}{3}R^{2}-G$,
respectively. It turns out that $S_{*}=S$ and $T_{*}=T$ is true
not only on the background ($H,q$) but also on the perturbative level
($\Phi,\Psi$): \begin{eqnarray}
G & = & -24H^{4}q+\ddot{\Phi}[24H^{2}]+\dot{\Phi}[48H^{3}(1-q)]+\dot{\Psi}[-24H^{3}]\nonumber \\
 &  & +a^{-2}\nabla^{2}\Psi[-8H^{2}]+a^{-2}\nabla^{2}\Phi[16H^{2}q]+\Psi[96H^{4}q].\end{eqnarray}
As a check one finds $\int d^{4}x\,\sqrt{|g+\delta g|}(G+\delta G)=(\text{surface term})$
with $G$ and $g$ unperturbed in this equation.

For solving the Einstein field equations $G_{\,\, b}^{a}=-8\pi G\cdot T_{\,\, b}^{a}$
we need the perturbed components of the Einstein tensor $G_{\,\, b}^{a}=R_{\,\, b}^{a}-\frac{1}{2}g_{\,\, b}^{a}R$,\begin{eqnarray}
G_{\,\,0}^{0} & = & -3H^{2}-6H\dot{\Phi}+6H^{2}\Psi+2a^{-2}\nabla^{2}\Phi\\
G_{\,\, j}^{i} & = & -a^{-2}(\Psi_{,j}^{\,\,,i}+\Phi_{,j}^{\,\,,i})+\delta_{\,\, j}^{i}\left[-\frac{1}{2}R-a^{-2}(1-2\Phi)(\delta_{ik}\text{-terms from }R_{ik})\right]\\
G_{\,\, k}^{0} & = & 2\dot{\Phi}_{,k}-2H\Psi_{,k}.\end{eqnarray}
Moreover, we apply a spatial Fourier decomposition for the perturbation
variables,\[
\Psi(t,\vec{x})=\int\frac{d^{3}k}{(2\pi)^{3}}\exp(-i\vec{k}\vec{x})\Psi(t,\vec{k})\]
with $k_{i}=\hat{k}_{i}\cdot k,\,\,\,\hat{k}_{j}\hat{k}^{j}=1,\,\,\, k=|\vec{k}|$
and $\Psi_{,i}=-ik_{i}\Psi,\,\,\,\nabla^{2}\Psi=-k^{2}\Psi$. Finally,
the longitudinal traceless part projection operator $P_{i}^{\,\, j}:=(\hat{k}_{i}\hat{k}^{j}-\frac{1}{3}\delta_{i}^{j})$
is introduced. It has the properties $P_{i}^{\,\, j}\delta_{j}^{i}=0$
and \[
P_{i}^{\,\, j}G_{\,\, j}^{i}=a^{-2}\left(k^{i}k_{j}\hat{k}_{i}\hat{k}^{j}-\frac{1}{3}k^{2}\right)(\Psi+\Phi)=\frac{2}{3}\left(\frac{k}{a}\right)^{2}(\Psi+\Phi)=(8\pi G)P_{i}^{\,\, j}T_{\,\, j}^{i},\]
when applied to the spatial components of the Einstein equation. If
there is no matter component with anisotropic stress, the right-hand
side vanishes and therefore~$\Psi=-\Phi$. The matter content discussed
in this paper fulfils this condition, however, we will keep the general
formulas for future work.

\subsection{Varying CC term}

The energy-momentum tensor of the cosmological term $T_{\Lambda\,\,\nu}^{\,\,\mu}=\rho_{\Lambda}\, g_{\,\,\nu}^{\mu}$
is the product of the metric~$g_{\,\,\nu}^{\mu}=\delta_{\,\,\nu}^{\mu}$
and the vacuum energy density $\rho_{\Lambda}=\rho_{\Lambda}^{i}+\beta/f$
as given by Eq.~(\ref{eq:rhoL-Def}). Since $\rho_{\Lambda}^{i}$
and $\beta$ are constants, only the function~$f=B/R$ has perturbations:\[
\delta T_{\Lambda\,\,\nu}^{\,\,\mu}=g_{\,\,\nu}^{\mu}\,\delta\rho_{\Lambda},\,\,\,\,\delta\rho_{\Lambda}=\beta\,\delta\left(\frac{1}{f}\right)=-\beta\,\frac{\delta f}{f^{2}}.\]
Applying the results from Sec.~\ref{sub:Pert-Grav} for determining
$\delta f=\delta(B/R)$ we find\begin{equation}
\delta\rho_{\Lambda}=\frac{\beta R}{B}\cdot\frac{N}{BR}\stackrel{\text{Relax}}{=}-\mathcal{O}(\rho_{\Lambda}^{i})\frac{N}{BR}\label{eq:drL-Relax}\end{equation}
with the numerator~$N$ given by\begin{equation}
N:=-(R^{2}+S)\delta R+R\delta S-y(R^{2}T\delta R+R^{3}\delta T).\label{eq:N}\end{equation}
The label {}``Relax'' means being in the relaxation regime, where~$|\rho_{\Lambda}|=|\rho_{\Lambda}^{i}+\beta R/B|\ll|\rho_{\Lambda}^{i}|$
and thus $\beta R/B\approx-\rho_{\Lambda}^{i}$. Furthermore, in this
regime, the term~$BR$ can be expressed in a simple form. First,
the constant~$\beta=-\rho_{\Lambda}^{i}9H_{*}^{2}$ is determined
by $\rho_{\Lambda}\approx0$ in the final de~Sitter era ($q=-1$)
with Hubble rate~$H_{*}$. Next, we plug $\beta$ back into the relaxation
relation $\rho_{\Lambda}^{i}+\beta R/B\approx0$ and obtain\begin{equation}
BR\stackrel{\text{Relax}}{=}9H_{*}^{2}R^{2}.\label{eq:BR}\end{equation}
Note that $BR\neq0$ even in the radiation regime, where $R^{2}\approx1/(2y)$
was found in Sec.~\ref{sub:Rad-Tracking}.

\subsection{Conservation equations\label{sub:Pert-Conservation}}

The energy-momentum tensor for the interacting $\Lambda$-X-component
with dust-like EOS~$\omega_{X}=0$ of the dark/X matter component
is given by\[
T_{\,\, b}^{a}=\rho_{X}u^{a}u_{b}+\rho_{\Lambda}g_{\,\, b}^{a},\]
which is covariantly conserved according to the Bianchi identity.
The perturbed versions of the energy densities and $4$-velocity $u^{a}$
are introduced by \[
\rho_{i}\rightarrow\rho_{i}(t)+\delta\rho_{i}(t,\vec{x}),\,\,\, i=X,\Lambda;\,\,\, u^{a}\rightarrow u^{a}(t)+\delta u^{a}(t,\vec{x}),\]
where in our coordinates \[
u^{a}=\delta_{0}^{a},\,\,\,\delta u^{0}=-\Psi=-\delta u_{0},\,\,\,\delta u^{j}=\frac{v^{j}}{a},\,\,\,\delta u_{j}=-av^{j}\]
with~$v^{j}$ being the $3$-velocity perturbation. From the $b=0$
component of the Bianchi identity~$T_{\,\, b;a}^{a}=0$ we find\[
T_{\,\,0;a}^{a}=[\dot{\rho}_{X}+\dot{\rho}_{\Lambda}+3H\rho_{X}]+\dot{\delta\rho}_{\Lambda}+\dot{\delta\rho}_{X}+3\dot{\Phi}\rho_{X}+3H\delta\rho_{X}+a^{-1}\rho_{X}v_{\,\,,j}^{j}=0,\]
where the vanishing of the zero-order term in square brackets is just
the conservation equation~(\ref{eq:L-X-Conservation}). Respectively,
the $b=j$ components lead to \begin{eqnarray}
T_{\,\,\,\,;a}^{aj} & = & -a^{-2}\delta\rho_{\Lambda,j}+a^{-2}\rho_{X}\Psi_{,j}+(\dot{\rho}_{X}+4H\rho_{X})\frac{v^{j}}{a}+\rho_{X}\frac{\dot{v}^{j}}{a}\\
 & = & -a^{-2}\delta\rho_{\Lambda,j}+a^{-2}\rho_{X}\Psi_{,j}+\frac{d}{dt}\left(\rho_{X}\frac{v^{j}}{a}\right)+5H\rho_{X}\frac{v^{j}}{a}=0.\label{eq:Cons-Bianchi-2}\end{eqnarray}
Since we assume irrotational matter the relations $v^{j}=(k^{j}/k)\, v$
and $v_{\,\,,j}^{j}=-ikv$ hold in Fourier space.

\subsection{Perturbations of radiation}

The radiation distribution function can be described by the moments~$\Theta_{a=0,1,2,\dots}$,
where $\Theta_{0}$ represents the monopole, $\Theta_{1}$ the dipole
and so on, cf.~Ref.~\cite{ModernCosmo}. For the purpose of this
paper we can ignore higher moments. The perturbed energy-momentum
tensor is given by \[
\delta\rho_{r}=\delta T_{\,\,0}^{0}=4\rho_{r}\Theta_{0},\,\,\,\delta T_{\,\, j}^{0}=4i\rho_{r}\Theta_{1}k^{j}\frac{a}{k}.\]
The evolution equations for $\Theta_{0,1}$ read\begin{eqnarray}
\dot{\Theta}_{0} & = & -\dot{\Phi}+\frac{k}{a}\Theta_{1},\label{eq:RTheta0-dot}\\
\dot{\Theta}_{1} & = & -\frac{1}{3}\frac{k}{a}(\Psi+\Theta_{0})+\frac{2}{3}\frac{k}{a}\Theta_{2}.\end{eqnarray}
This set can be decoupled to calculate the evolution of $\Theta_{0,1}$
when $\Phi,\Psi$ are known:\begin{eqnarray}
(\ddot{\Theta}_{0}+\ddot{\Phi})+H(\dot{\Theta}_{0}+\dot{\Phi})+\frac{1}{3}\left(\frac{k}{a}\right)^{2}(\Theta_{0}+\Psi) & = & 0,\label{eq:Theta0-ODE}\\
\ddot{\Theta}_{1}+H\dot{\Theta}_{1}+\frac{1}{3}\left(\frac{k}{a}\right)\left(\dot{\Psi}-\dot{\Phi}+\left(\frac{k}{a}\right)\Theta_{1}\right) & = & 0.\label{eq:Theta1-ODE}\end{eqnarray}
The quadrupole term~$\Theta_{2}$ has been neglected in these equations.

\subsection{Einstein equations for the perturbations}

From Sec.~\ref{sub:Pert-Grav} we will use three components of $\delta G_{\,\, b}^{a}=8\pi G\,\delta T_{\,\, b}^{a}$
in Fourier space:\begin{eqnarray}
6H\dot{\Phi}-6H^{2}\Psi+2\left(\frac{k}{a}\right)^{2}\Phi & = & 8\pi G(\delta\rho_{\Lambda}+\delta\rho_{X}+\delta\rho_{r}),\label{eq:PEinstein1}\\
-i2k^{j}(\dot{\Phi}-H\Psi) & = & 8\pi G\left(\rho_{X}av^{j}-4i\rho_{r}\Theta_{1}\frac{k^{j}}{k}a\right),\label{eq:PEinstein2-k}\\
\frac{1}{8\pi G}\frac{2}{3}\left(\frac{k}{a}\right)^{2}(\Psi+\Phi) & = & P_{i}^{\,\, j}T_{\,\, j}^{i}=-\frac{8}{3}\rho_{r}\Theta_{2}=\text{anisotropic part.}\label{eq:PEinstein3}\end{eqnarray}
By contracting with~$k^{j}$, the second equation can be rewritten
in the more useful form\begin{equation}
\rho_{X}v\frac{k}{a}=4i\rho_{r}\Theta_{1}\frac{k}{a}-\frac{2i}{8\pi G}\left(\frac{k}{a}\right)^{2}(\dot{\Phi}-H\Psi).\label{eq:PEinstein2}\end{equation}
Combining these equations with the second of the following conservation
equations from Sec.~\ref{sub:Pert-Conservation},\begin{eqnarray}
\dot{\delta\rho}_{\Lambda}+\dot{\delta\rho}_{X}+3\dot{\Phi}\rho_{X}+3H\delta\rho_{X}-i\rho_{X}v\frac{k}{a} & = & 0,\label{eq:TBianchi1}\\
\delta\rho_{\Lambda}\left(\frac{k}{a}\right)^{2}-\rho_{X}\Psi\left(\frac{k}{a}\right)^{2}-i\frac{d}{dt}\left(\rho_{X}v\frac{k}{a}\right)-i5H\left(\rho_{X}v\frac{k}{a}\right) & = & 0,\label{eq:TBianchi2}\end{eqnarray}
we find\begin{eqnarray}
\delta\rho_{\Lambda} & = & \rho_{X}\Psi+\frac{4}{3}\rho_{r}(\Theta_{0}+\Psi-2\Theta_{2})+\frac{2}{8\pi G}[\ddot{\Phi}+3H\dot{\Phi}-H\dot{\Psi}-\dot{H}\Psi-3H^{2}\Psi],\label{eq:deltaL}\\
\delta\rho_{X} & = & -\rho_{X}\Psi-\frac{4}{3}\rho_{r}(4\Theta_{0}+\Psi-2\Theta_{2})-\frac{2}{8\pi G}\left[\ddot{\Phi}-H\dot{\Psi}-\dot{H}\Psi-\left(\frac{k}{a}\right)^{2}\Phi\right].\label{eq:deltaX}\end{eqnarray}
Alternatively, the first conservation equation~(\ref{eq:TBianchi1})
leads to\begin{eqnarray}
\delta\rho_{X} & = & -\frac{4}{3}(4\rho_{r}\Theta_{0})-\frac{2}{8\pi G}\left[\ddot{\Phi}-H\dot{\Psi}-2\dot{H}\Psi+\frac{1}{3}\left(\frac{k}{a}\right)^{2}(\Psi-2\Phi)\right],\label{eq:dX+4/3dr}\\
\delta\rho_{\Lambda}+\frac{1}{4}\delta\rho_{X} & = & -2\Psi(\rho_{\Lambda}+\frac{1}{4}\rho_{X})+\frac{3}{2}\frac{1}{8\pi G}\left[\ddot{\Phi}+4H\dot{\Phi}-H\dot{\Psi}+\frac{1}{3}\left(\frac{k}{a}\right)^{2}(\Psi+2\Phi)\right],\label{eq:dL+dX/4}\end{eqnarray}
where the second line follows from the first and Eq.~(\ref{eq:PEinstein1}),
respectively.

\subsection{Evolution of the gravitational potentials and perturbations}

\subsubsection{Gravitational potentials\label{sub:Gravitational-potentials}}

Let us estimate the order of magnitude of the vacuum energy perturbation~$\delta\rho_{\Lambda}$.
Since by definition all perturbation variables $\Phi,\Psi,\Theta_{0},\dots$
are expected to be small at some initial time, we obtain a good estimate
of the magnitude of $\delta\rho_{\Lambda}$ by using for instance
Eq.~(\ref{eq:deltaL}), thus $\delta\rho_{\Lambda}=\mathcal{O}(\rho_{c}\Phi)$
with $\rho_{c}=3H^{2}/(8\pi G)$. On the other hand, we know that
in the relaxation regime~$|\rho_{\Lambda}|=|\rho_{\Lambda}^{i}+\frac{\beta}{f}|\ll|\rho_{\Lambda}^{i}|$,
which leads via~(\ref{eq:drL-Relax}) to\begin{eqnarray}
\mathcal{O}(\rho_{c}\Phi) & = & \delta\rho_{\Lambda}=-\beta\frac{\delta f}{f^{2}}=-\rho_{\Lambda}^{i}\frac{\delta f}{f}\\
\Longrightarrow\,\,\frac{\delta f}{f} & = & \frac{N}{BR}=-\frac{\delta\rho_{\Lambda}}{\rho_{\Lambda}^{i}}=\mathcal{O}\left(\frac{\rho_{c}\Phi}{\rho_{\Lambda}^{i}}\right).\label{eq:bEQzero}\end{eqnarray}
Due to the suppression by the large initial term~$\rho_{\Lambda}^{i}$,
the righthand side of the last equation is extremely small, implying
$|N/(BR)|\ll1$. In addition, $BR\neq0$ from Eq.~(\ref{eq:BR})
is finite, and we conclude that the evolution of the gravitational
potentials~$\Phi,\Psi$ can be determined in excellent approximation
by solving the equation $N=0$ (instead of~$N/(BR)=\epsilon\ll1$),
where $N$ is given in Eq.~(\ref{eq:N}). Note, however, that neither~$N$
nor~$\delta\rho_{\Lambda}$ vanish as stated explicitely in Eq.~(\ref{eq:bEQzero}).
In the following sections, we will use {}``$N=0$'' as a shortcut
for analysing the gravitational potentials, but not for calculating~$\delta\rho_{\Lambda}$
directly via Eq.~(\ref{eq:drL-Relax}). Instead,~$\delta\rho_{\Lambda}$
will be determined by one of the Eqs.~(\ref{eq:deltaL},\ref{eq:deltaX},\ref{eq:dX+4/3dr},\ref{eq:dL+dX/4})
after the evolution of~$\Phi,\Psi$ has been found from solving {}``$N=0$''.

\subsubsection{Matter era}

In the matter epoch the deceleration variable $q=\frac{1}{2}$ by
construction and consequently $H=2/(3t)\ll H_{eq}$, $R=3H^{2}$,
$S=9H^{4}$, $T=15H^{4}$. This means that in Eq.~(\ref{eq:N}) we
can neglect the part proportional to $y=H_{eq}^{-4}$ and find\begin{eqnarray}
N & = & -2R^{2}\delta R+R\delta S=-18H^{4}[6\ddot{\Phi}+18H\dot{\Phi}-6H\dot{\Psi}-2a^{-2}\nabla^{2}(\Psi+\Phi)]\\
 & = & -6\cdot18H^{4}[\ddot{\Phi}+4H\dot{\Phi}]\,\,\,\,\,(\text{with }\Psi=-\Phi).\end{eqnarray}
As discussed in Sec.~\ref{sub:Gravitational-potentials}, now we
use {}``$N=0$'' as a shortcut for determining the gravitational
potential~$\Phi$. Accordingly, we solve $\ddot{\Phi}+4H\dot{\Phi}=0$,
which has the solutions~$\Phi\propto t^{-5/3}$ and~$\Phi=\text{const.}$.
While the decaying mode will die out, the constant gravitational potential
is a solution in $\Lambda$CDM, too. To calculate the perturbations
of $\Lambda$, we cannot use directly the expression $\delta\rho_{\Lambda}=\beta\frac{R}{B}\cdot\frac{N}{BR}$
because we already required $N=0$ to determine~$\Phi$. However,
Eq.~(\ref{eq:deltaL}) is suitable for this purpose and we obtain
\[
\delta\rho_{\Lambda}=-\rho_{X}\Phi+\frac{4}{3}\rho_{r}(\Theta_{0}-\Phi)+\frac{3H^{2}}{8\pi G}\Phi=\Phi\left(\rho_{c}-\rho_{X}-\frac{4}{3}\rho_{r}\right)+\frac{4}{3}\rho_{r}\Theta_{0}=4\rho_{\Lambda}\Theta_{0},\]
where we used in the last step the tracking property $\rho_{r}=3\rho_{\Lambda}$
from Eq.~(\ref{eq:Mat-3rL--rR}). In the matter era, Eq.~(\ref{eq:Theta0-ODE})
for the radiation monopole has the solution\[
\Theta_{0}=\Phi+c_{1}\cos\left(\frac{2}{\sqrt{3}}\frac{k}{aH}\right)+c_{2}\sin\left(\frac{2}{\sqrt{3}}\frac{k}{aH}\right),\]
which implies that the $\Lambda$~density contrast $\delta\rho_{\Lambda}/\rho_{\Lambda}\simeq4\Phi$
is constant in magnitude on average. Finally, Eq.~(\ref{eq:deltaX})
determines the evolution of $\delta\rho_{X}$,\begin{equation}
\delta\rho_{X}=-\frac{16}{3}\rho_{r}\Theta_{0}+\left[2\rho_{c}+\frac{2}{8\pi G}\left(\frac{k}{a}\right)^{2}\right]\Phi\approx\rho_{X}\Phi\left[2+\frac{2k^{2}}{3H_{0}^{2}\Omega_{X0}}\cdot a\right],\label{eq:Mat-deltaX}\end{equation}
where we used $\rho_{c}\approx\rho_{X}\approx\Omega_{X0}\rho_{c}^{0}a^{-3}\gg\rho_{r}$
in the last step. We conclude that for sub-horizon ($k/(aH)\gg1$)
modes the X/dark matter density contrast\begin{equation}
D:=\frac{\delta\rho_{X}}{\rho_{X}}\propto a\label{eq:Mat-Xcontrast}\end{equation}
grows linearly with the scale factor~$a$ providing the basis of
structure formation. This is the same behaviour as in the $\Lambda$CDM
model. It should be clear that once $D$ becomes of order $1$, the
perturbative analysis breaks down.

Let us confirm Eq.~(\ref{eq:Mat-Xcontrast}) by directly solving
the differential equation for $\delta\rho_{X}$ or respectively $D$.
This is possible for negligible radiation and vacuum perturbations
($\delta\rho_{r,\Lambda}\ll\delta\rho_{X}$) in the sub-horizon limit,
where $k/(aH)\gg1$ and $|\dot{\Phi}|\ll|v\, k/a|$. Accordingly,
we find the Poisson equation\[
\left(\frac{k}{a}\right)^{2}\Phi=\frac{8\pi G}{2}\delta\rho_{X}=\frac{8\pi G}{2}\rho_{X}D\]
from Eqs.~(\ref{eq:PEinstein1}) and (\ref{eq:PEinstein2}). Furthermore,
we take the time-derivative of Eq.~(\ref{eq:TBianchi1}) and remove
all terms involving the velocity $v$ with the help of Eqs.~(\ref{eq:TBianchi2},\ref{eq:TBianchi1}).
Also, we introduce the function~$Q:=-\dot{\rho}_{\Lambda}/\rho_{X}$,
which occurs in the relation $\dot{\delta\rho}_{X}=\rho_{X}(\dot{D}+(Q-3H)D)$
as a result of the conservation equation~(\ref{eq:L-X-Conservation}).
Finally, with the Poisson equation $\Phi$ is eliminated from Eq.~(\ref{eq:TBianchi2})
and we obtain\begin{equation}
\ddot{D}+2(H+Q)\dot{D}+\left(\dot{Q}+Q^{2}+2HQ-4\pi G\rho_{X}\right)D=0.\label{eq:Mat-D-ODE}\end{equation}
In the matter era, the tracking relation (\ref{eq:Mat-3rL--rR}) implies
$Q=\frac{4}{3}H\rho_{r}/\rho_{X}$ and Eq.~(\ref{eq:Friedmann-H})
becomes \[
4\pi G\rho_{X}=\frac{3}{2}H^{2}\left(1+\frac{4}{3}\frac{\rho_{r}}{\rho_{X}}\right)^{-1}=\frac{3}{2}H^{2}\left(1-\frac{4}{3}\frac{\rho_{r}}{\rho_{X}}+\mathcal{O}\left(\frac{\rho_{r}}{\rho_{X}}\right)^{2}\right).\]
Neglecting the second-order correction in the last step we find for
$D$ two solutions from Eq.~(\ref{eq:Mat-D-ODE}),\[
D=c_{1}\left(x+\frac{8}{3}+\frac{80}{9}x^{-1}\right)+c_{2}\left(x^{-\frac{3}{2}}+\mathcal{O}\left(x^{-\frac{5}{2}}\right)\right),\,\,\,\,\, x:=a\cdot\frac{\rho_{X0}}{\rho_{r0}},\]
where $\rho_{r0}\ll\rho_{X0}$ and $c_{1,2}$ are integration constants.
In the limit~$x\gg1$ this result confirms Eq.~(\ref{eq:Mat-Xcontrast}).
Moreover, it agrees with the corresponding discussion about the density
contrast in Ref.~\cite{Pert-LXCDM-1}.

\subsubsection{De Sitter era}

In the final de~Sitter regime we have $q=-1$, $H=H_{*}=\text{const.}$,
$R=12H^{2}$, $S=36H^{4}$ and $T=24H^{4}$, respectively, and the
terms proportional to $y=H_{eq}^{-4}$ can be neglected. Also here,
the arguments from Sec.~\ref{sub:Gravitational-potentials} imply
that solving~$N=0$ yields the evolution of~$\Phi$ with $N$ having
the form\begin{eqnarray}
N & = & -(R^{2}+S)\delta R+R\delta S=-18\cdot36H^{4}\left[\ddot{\Phi}+4H\dot{\Phi}-H\dot{\Psi}-\frac{1}{3}a^{-2}\nabla^{2}\Psi-\frac{2}{3}a^{-2}\nabla^{2}\Phi-4H\Psi\right]\nonumber \\
 & = & -18\cdot36H^{4}\left[\ddot{\Phi}+5H\dot{\Phi}-\frac{1}{3}a^{-2}\nabla^{2}\Phi+4H^{2}\Phi\right]\,\,\,\,\,(\text{with }\Psi=-\Phi).\end{eqnarray}
In Fourier space this means\begin{equation}
\ddot{\Phi}+5H\dot{\Phi}+\frac{1}{3}\left(\frac{k}{a}\right)^{2}\Phi+4H^{2}\Phi=0,\label{eq:dS-Phi-Eq}\end{equation}
which has two solutions\begin{eqnarray}
\Phi_{1} & = & x^{2}(\sin x+x^{-1}\cos x)=x+\frac{1}{2}x^{3}+\mathcal{O}(x^{5})\\
\Phi_{2} & = & x^{2}(x^{-1}\sin x-\cos x)=\frac{1}{3}x^{4}+\mathcal{O}(x^{6}),\end{eqnarray}
where \[
x:=\frac{k}{\sqrt{3}aH}.\]
Since~$H$ is constant, both modes~$\Phi_{1,2}$ are decaying for
all wavenumbers~$k$. With $\dot{x}=-Hx$ the time derivatives read\[
\dot{\Phi}_{1}=-H(\Phi_{1}+x^{3}\cos x),\,\,\,\,\dot{\Phi}_{2}=-H(\Phi_{2}+x^{3}\sin x).\]
The density perturbations of~$\Lambda$ and~X follow from Eqs.~(\ref{eq:deltaL},\ref{eq:deltaX},\ref{eq:dS-Phi-Eq})
and the solutions~$\Phi_{1,2}$ of the gravitational potentials:\begin{eqnarray}
\delta\rho_{\Lambda} & = & -\rho_{X}\Phi_{1}+\frac{3H^{2}}{8\pi G}\left(\frac{2}{3}x^{3}\cos x-\frac{2}{3}x^{2}\Phi_{1}\right),\\
\delta\rho_{X} & = & \rho_{X}\Phi_{1}+\frac{3H^{2}}{8\pi G}\left(-\frac{8}{3}x^{3}\cos x+\frac{8}{3}x^{2}\Phi_{1}\right).\end{eqnarray}
Here, the corresponding results for~$\Phi=\Phi_{2}$ can be obtained
by replacing~$\cos x$ by $\sin x$. The zero-order Einstein equation
requires~$\rho_{X}=0$ leading to~$\delta\rho_{X}=-4\delta\rho_{\Lambda}$
and\[
\left.\frac{\delta\rho_{\Lambda}}{\rho_{\Lambda}}\right|_{\Phi=\Phi_{1}}=-\frac{2}{3}(x^{4}\sin x),\,\,\,\,\left.\frac{\delta\rho_{\Lambda}}{\rho_{\Lambda}}\right|_{\Phi=\Phi_{2}}=\frac{2}{3}(x^{4}\cos x).\]
 Therefore, the $\Lambda$ density contrast is decreasing and no perturbative
instabilities are to be expected in the future.

\subsubsection{Radiation era\label{sub:Radiation-era}}

In the radiation epoch the expansion variables read $q=1$, $S=12H^{4}$
and $T=24H^{4}$, respectively. Applying the considerations from Sec.~\ref{sub:Rad-Tracking}
we find in the relaxation regime\begin{equation}
R\approx\frac{1}{\sqrt{2y}}\sim H_{eq}^{2}\ll H^{2},\,\,\,\, BR=9H_{*}^{2}R^{2}\approx\frac{9H_{*}^{2}}{2y}.\end{equation}
Moreover, from Eq.~(\ref{eq:N}) we obtain \begin{eqnarray}
N & = & -(R^{2}+S)\delta R+R\delta S-y(R^{2}T\delta R+R^{3}\delta T)\\
 & = & -\left(R^{2}+S+\frac{1}{2}T\right)\delta R+R\delta S-\frac{1}{2}R\delta T\\
 & \approx & -24H^{4}\delta R,\end{eqnarray}
where terms proportional to $H_{eq}\ll H$ have been neglected. Finally,
the shortcut {}``$N=0$'' from Sec.~\ref{sub:Gravitational-potentials}
implies (with~$\Psi=-\Phi$) the equation\begin{equation}
0=\delta R=6\left(\ddot{\Phi}+5H\dot{\Phi}+\frac{1}{3}\left(\frac{k}{a}\right)^{2}\Phi\right),\label{eq:Rad-Phi-EOM}\end{equation}
which is relevant for the evolution of $\Phi$. The corresponding
solutions read\begin{eqnarray}
\Phi_{1} & = & 3\Phi_{0}x^{-2}(\sin x+x^{-1}\cos x)=\Phi_{0}\left(3x^{-3}+\frac{3}{2}x^{-1}+\mathcal{O}(x)\right),\\
\Phi_{2} & = & 3\Phi_{0}x^{-2}(x^{-1}\sin x-\cos x)=\Phi_{0}\left(1-\frac{1}{10}x^{2}+\mathcal{O}(x^{4})\right),\end{eqnarray}
and they are equal to the solutions in $\Lambda$CDM, where radiation
determines~$\Phi$ alone. However, notice the remark at the end of
this section. With~$a=(t/t_{0})^{1/2}$, $H=1/(2t)$ and \[
x:=\sqrt{\frac{k^{2}}{3a^{2}H^{2}}}\propto a\propto\sqrt{t},\,\,\,\,\dot{x}=Hx\]
we find in the early-time limit~$t,a\rightarrow0$\[
\Phi_{1}\sim t^{-\frac{3}{2}},\,\,\,\Phi_{2}\rightarrow\Phi_{0}=\text{const.}\]
Therefore, only~$\Phi_{2}$ has a regular early-time behaviour $\Phi(t\rightarrow0)=\Phi_{0}$.
Using this solution from now on, one can solve Eqs.~(\ref{eq:Theta0-ODE})
and~(\ref{eq:Theta1-ODE}) for the radiation potentials $\Theta_{0,1}$,
which are found to be (with $\Psi=-\Phi$)\begin{eqnarray}
\Theta_{0}(x) & = & c_{1}\cos x+c_{2}\sin x-\Phi(x)+3\Phi_{0}\frac{\sin x}{x},\label{eq:Theta0-Sol}\\
\Theta_{1}(x) & = & \frac{1}{\sqrt{3}}\left(-c_{1}\sin x+c_{2}\cos x-x\Phi(x)\right).\label{eq:Theta1-Sol}\end{eqnarray}
However, note these results assume only two non-vanishing multipole
modes $\Theta_{0,1}$ and no interaction with baryons.

The evolutions of the $\Lambda$ and X components follow directly
from Eqs.~(\ref{eq:dX+4/3dr}) and~(\ref{eq:dL+dX/4}), respectively.
We find\[
\delta\rho_{X}=4\Theta_{0}\rho_{X}-\frac{2}{3}\rho_{c}\left((8c_{1}+12\Phi_{0})+8c_{2}\sin x\right)=-4\delta\rho_{\Lambda}.\]
In the case of $\Lambda$CDM initial conditions ($c_{1}=-\frac{3}{2}\Phi_{0},\,\, c_{2}=0$,
see Sec.~\ref{sub:Initial-conditions}), we obtain during the tracking
regime\[
\frac{\delta\rho_{X}}{\rho_{X}}=4\Theta_{0}=\frac{\delta\rho_{\Lambda}}{\rho_{\Lambda}}=\frac{\delta\rho_{r}}{\rho_{r}},\]
indicating that the perturbations of $\Lambda$, X and radiation evolve
in the same manner.

A remark is in order. When using Eqs.~(\ref{eq:dX+4/3dr}) and~(\ref{eq:dL+dX/4})
in the $\Lambda$CDM model, one must not forget the small but relevant
change in the evolution equation~(\ref{eq:Rad-Phi-EOM}) for $\Phi$
due to the presence of dust matter. Otherwise, one would erroneously
infer that the tracking relation~$\delta\rho_{X}+4\delta\rho_{\Lambda}=0$
is valid also in $\Lambda$CDM, where $\delta\rho_{\Lambda}\equiv0$.
For this case, it is better to work with Eqs.~(\ref{eq:TBianchi1})
and~(\ref{eq:TBianchi2}).

\subsection{Initial conditions\label{sub:Initial-conditions}}

We determine the initial conditions for $\delta\rho_{r}$, $\delta\rho_{\Lambda}$
and $\delta\rho_{x}$ in terms of the primordial value~$\Phi_{0}(k)$
of the gravitational potential (which we assume to be fixed by inflation).
Since all perturbation modes relevant today were super-Hubble modes
at that time, we will use the small wavenumber $x=k/(\sqrt{3}aH)\ll1$
as expansion parameter. Accordingly, we set $k=0$ for the initial
conditions of $\delta\rho_{r,\Lambda,X}$ and keep the leading order
in $k$ for $\Theta_{1},v$. 

Let us start with the case in which the relaxation mechanism is not
active yet, i.e.\ radiation dominates the cosmological evolution
alone:\[
\rho_{r}\gg|\rho_{\Lambda,X}|,\,\rho_{\Lambda}=\rho_{\Lambda}^{i}=\text{const.},\,|\delta\rho_{\Lambda}|\ll\Phi\rho_{c},\,\dot{H}=-2H^{2},\,\Psi=-\Phi.\]
From the time derivative of Eq.~(\ref{eq:PEinstein1}) we find with
$k=0$ \begin{eqnarray}
6H\ddot{\Phi}-6H^{2}\dot{\Phi}-24H^{3}\Phi & = & 8\pi G\,4\rho_{r}(\dot{\Theta}_{0}-4H\Theta_{0})\\
 & = & -4H(6H\dot{\Phi}+6H^{2}\Phi)-12H^{2}\dot{\Phi},\end{eqnarray}
where we used~$\rho_{r}\approx\rho_{c}=3H^{2}/(8\pi G)$ and the
Eqs.~(\ref{eq:RTheta0-dot}) and~(\ref{eq:PEinstein1}). The resulting
equation~$\ddot{\Phi}+5H\dot{\Phi}=0$ has two solutions $\Phi\propto a^{-3}$
and $\Phi=\Phi_{0}=\text{const}$. The first one is decaying and will
die out quickly, thus we will consider only the constant mode. With
$\dot{\Phi}=0$ Eqs.~(\ref{eq:PEinstein1}) and~(\ref{eq:PEinstein2})
yield the initial conditions \[
\Theta_{0}^{i}=\frac{1}{2}\Phi_{0},\,\,\,\Theta_{1}^{i}=\frac{1}{6}\frac{k}{aH}\Phi_{0}\]
Moreover, the corresponding conditions for the X matter follow from
the Bianchi identity~(\ref{eq:TBianchi1}) with $k=0$, and respectively
from~(\ref{eq:TBianchi2}) with~$k\neq0$,\[
\delta\rho_{X}^{i}=\rho_{X}(-3\Phi_{0}+c_{3}),\,\,\, iv^{i}=\frac{1}{2}\frac{k}{aH}\Phi_{0}=3\Theta_{1}^{i},\]
where $c_{3}$ is a constant. Since the relaxation mechanism is not
active, the results above are identical to those in the $\Lambda$CDM
model as expected.

Now we discuss the case where the tracking regime ($\rho_{x}=-4\rho_{\Lambda}\propto\rho_{r}$)
already starts right after reheating or whatever was before the radiation
era, thus we cannot assume $\rho_{X,\Lambda}\ll\rho_{r}$. According
to Eqs.~(\ref{eq:dL+dX/4}) and~(\ref{eq:Rad-Phi-EOM}) the perturbations
of $\rho_{X,\Lambda}$ obey the tracking relation $\delta\rho_{X}+4\delta\rho_{\Lambda}=0$.
Then we solve the conservation equation~(\ref{eq:TBianchi1}) for
$\delta\rho_{X}$ in the $k\rightarrow0$ limit, which yields\[
\delta\rho_{X}^{i}=\rho_{X}(-4\Phi_{0}+c_{3}),\]
with a constant~$c_{3}$. Applying again~$\delta\rho_{\Lambda}=-\delta\rho_{X}/4$,
Eq.~(\ref{eq:TBianchi2}) can be solved for the initial X velocity
perturbation~$v$:\[
(iv)^{i}=\left(2\Phi_{0}-\frac{1}{4}c_{3}\right)\frac{k}{aH}.\]
The initial condition for the radiation monopole, $\Theta_{0}^{i}=c_{1}+2\Phi_{0}$,
follows from Eq.~(\ref{eq:Theta0-Sol}) in the limit $x\rightarrow0$,
whereas for $\Theta_{1}^{i}$ we evolve Eq.~(\ref{eq:Theta1-Sol})
up to the first power in $k$,\[
\Theta_{1}=-\frac{1}{\sqrt{3}}(c_{2}-c_{1}x-\Phi_{0}x)+\mathcal{O}(x^{2})=\frac{c_{2}}{\sqrt{3}}-\frac{x}{\sqrt{3}}(\Theta_{0}^{i}-\Phi_{0})+\mathcal{O}(x^{2}).\]
Since $(iv)^{i}\propto k$, Eq.~(\ref{eq:PEinstein2}) implies $\Theta_{1}^{i}\propto k$,
and we obtain from the last equation\[
c_{2}=0,\,\,\,\Theta_{1}^{i}=-\frac{x}{\sqrt{3}}(\Theta_{0}^{i}-\Phi_{0})=-\frac{k}{3aH}(\Theta_{0}^{i}-\Phi_{0}).\]
Finally, the constants~$c_{1}$ in~$\Theta_{0}^{i}$ and~$c_{3}$
in~$\delta\rho_{X}^{i}$ are related by~(\ref{eq:PEinstein1}),\[
2\rho_{c}\Phi_{0}=\frac{3}{4}\delta\rho_{X}^{i}+\delta\rho_{r}^{i}=\left(-3\Phi_{0}+\frac{3}{4}c_{3}\right)\rho_{X}+4\rho_{r}\Theta_{0}^{i}.\]
Therefore, the knowledge of either $c_{1}$, $c_{3}$ or $\delta\rho_{\Lambda}^{i}=-\delta\rho_{X}^{i}/4$
will uniquely fix all initial conditions. Remember that $\delta\rho_{\Lambda}^{i}$
can be calculated directly via Eq.~(\ref{eq:drL-Relax}) once an
inflation or reheating model determines the background evolution at
the initial time.

\subsection{Tensor perturbations\label{sub:Tensor-perturbations}}

According to the decomposition theorem, scalar, vector and tensor
perturbations of the metric~$g_{mn}$ evolve independently. Therefore,
we discuss in this section only tensor perturbations (gravitational
waves). For simplicity, let us consider a wave propagating in $z$-direction,
so that the wave amplitude is a function of the cosmic time~$t$
and $z$ only. The perturbed metric in linear order reads\[
g_{00}=1,\,\,\,\, g_{ij}=-a^{2}(t)(1+H_{ij}(t,z))=-a^{2}(t)\left(\begin{array}{ccc}
1+h_{1} & h_{2} & 0\\
h_{2} & 1-h_{1} & 0\\
0 & 0 & 1\end{array}\right),\]
where $h_{1,2}$ denote the amplitudes of the two polarisation modes,
and $H_{ij}$ is symmetric, traceless and divergenceless. Then the
Einstein tensor is given by \[
G_{\,\,0}^{0}=-3H^{2},\,\,\,\, G_{\,\,3}^{3}=-\left(2\frac{\ddot{a}}{a}+H^{2}\right),\,\,\,\, G_{\,\,2}^{1}=G_{\,\,1}^{2}=\frac{1}{2}(\ddot{h}_{2}+3H\dot{h}_{2}-a^{-2}\partial_{z}^{2}h_{2}),\]
\[
G_{\,\,1}^{1}=G_{\,\,3}^{3}+\frac{1}{2}(\ddot{h}_{1}+3H\dot{h}_{1}-a^{-2}\partial_{z}^{2}h_{1}),\,\,\,\, G_{\,\,2}^{2}=G_{\,\,3}^{3}-\frac{1}{2}(\ddot{h}_{1}+3H\dot{h}_{1}-a^{-2}\partial_{z}^{2}h_{1}).\]
Moreover, it turns out that the geometrical scalars $R=R_{\,\, a}^{a}$,
$S=R^{ab}R_{ab}$ and $T=R^{abcd}R_{abcd}$ do not depend on $h_{1,2}$
implying that the gravity wave propagation in the $\Lambda$XCDM relaxation
model is the same as in the $\Lambda$CDM model. Thus, from the Einstein
equations one obtains the standard equations for gravitational waves\[
G_{\,\,1}^{1}-G_{\,\,2}^{2}=\ddot{h}_{1}+3H\dot{h}_{1}-a^{-2}\partial_{z}^{2}h_{1}=-8\pi G(T_{\,\,1}^{1}-T_{\,\,2}^{2})=0,\]
\[
G_{\,\,1}^{2}=\frac{1}{2}(\ddot{h}_{2}+3H\dot{h}_{2}-a^{-2}\partial_{z}^{2}h_{2})=-8\pi GT_{\,\,1}^{2}=0,\]
where we assumed on the right-hand sides the absence of anisotropic
stress.

\section{Conclusions\label{sec:Conclusions}}

We discussed analytically the vacuum tracking properties and the evolution
of perturbations in the $\Lambda$XCDM relaxation model for a large
initial CC. Therefore, this work complements the background evolution
analysis in Ref.~\cite{Relax-LXCDM}. Concerning the tracking behaviour,
the vacuum energy density is found to be proportional to the radiation
energy density in the radiation era and in the matter era, too. In
the radiation epoch the energy densities of radiation, dark/X matter
and the vacuum scale in the same way, whereas deep in the matter era
the vacuum energy is always one third of the radiation energy. Thus,
in addition to avoiding the CC fine-tuning problem, this setup lets
the coincidence problem look less severe than in the $\Lambda$CDM
model.

Apart from the interesting tracking behaviour, we also analysed linear
perturbations in this model. We found that their evolution has many
similarities with the corresponding evolution in the $\Lambda$CDM
model. For the cases studied in this work we did not find any hints
of instability apart from the growing dark/X matter modes during the
matter era, which provide the seed for structure formation. As a result,
the perturbations are well-behaved according to the analytical discussion,
but at the same time, we did not find a new observational signature
of the relaxation model. However, deviations from $\Lambda$CDM are
expected to appear in the transition regions from radiation to matter
domination and from the matter era into the final de~Sitter stage.
In Ref.~\cite{Relax-LXCDM} several differences in the background
evolution have been discussed already, and we expect that the scale
factor evolution is the main source for observational signatures for
two reasons. First, neither the baryon-photon plasma nor neutrinos
couple to the dark sector in our model, and second, as a result of
this work, the growth of dark/X matter modes in the matter era is
more or less identical to $\Lambda$CDM. In summary, we found that
the $\Lambda$XCDM relaxation model works well in the perturbations
sector. When implementing the relaxation mechanism in a modified gravity
framework, we expect to encounter more instabilities because new degrees
of freedom are available. This subject is under investigation.

\subsection*{Acknowledgements}

I would like to thank Joan Solà and Hrvoje Štefančić for useful discussions
and comments.\\
This work has been supported in part by MEC and FEDER under project
FPA2007-66665, by the Spanish Consolider-Ingenio 2010 program CPAN
CSD2007-00042 and by DIUE/CUR Generalitat de Catalunya under project
2009SGR502.

\end{document}